# Upwind exciton-polariton propagation in hybrid photonic crystals.

Andrea D'Andrea and Norberto Tomassini

*Istituto dei Sistemi Complessi, CNR, Italy*
*Area della Ricerca di Roma 1 (Montelibretti)*
*(01-10-2016)*

In this work we study theoretically the axial spectral asymmetry of a 1D periodic multilayer systems composed by hybrid two-layers (isotropic/anisotropic) and for photon energies close to the electronic energy gaps of semiconductors (excitons). The non-normal optical properties of these resonant non-magnetic photonic crystals, where linear and quadratic spatial dispersion effects are both present, will be computed in the framework of exciton-polariton by self-consistent Maxwell-Schroedinger equation solutions in the effective mass approximation. The tailoring of hybrid crystal, where exciton-polariton unidirectional oblique propagation is observed, will be computed by implementing a simple two-layers "minimum model".

In the recent literature many new optical effects in hybrid (isotropic/anisotropic) photonic crystals are studied, namely: negative refraction, oblique frozen modes, giant transmission, absorption suppression and optical unidirectional propagation are among the most interesting ones [1-7]. Moreover, all the symmetry conditions, necessary in order to observe the former optical effects, are fully discussed. Recently, exciton-polariton, a composite particle coming from strong radiation-matter interaction [8], demonstrates a wide range of new phenomena in photonic crystal resonators: optical spin-hole effects, magnetic dipole enhancement, Bose-Einstein condensation at room temperature, radiative topological states [9-13].

In this letter we describe a novel effect involving 1D periodic hybrid meta-material that can transmit exciton-polariton as an unidirectional oblique wave. Let us consider a simple elementary cell of d-periodicity along z-axis, composed by two-layers (isotropic and anisotropic) of equal thicknesses; the dispersion curves, at normal incidence, are given by $\hbar\omega(K)$, where K is the Bloch wavevector of 1D lattice. The polariton propagation, at non-normal incidence, is described by a 3D Fourier transform for in-plane wavevector $\vec{K}_{//} \neq 0$ and total dispersion curves: $\hbar\omega(\vec{K}_{//}, K)$. By choosing (x,z) as plane of incidence, the in-plane wave vector becomes $\vec{K}_{//} = (K_x, 0, 0)$, where $K_x = \frac{\omega}{c}\sqrt{\varepsilon_b}\sin(\alpha)$ and $\alpha$ is the angle between propagation direction and z-axis, therefore a 2D Fourier transform, with dispersion curves $\hbar\omega(K_x, K)$ or $\hbar\omega_\alpha(K)$, is obtained, and for oblique propagation the components of group velocity are: $\vec{v}_g = \left(\frac{c}{\sqrt{\varepsilon_b}\sin(\alpha)}, 0, \frac{\partial\omega_\alpha(K)}{\partial K}\right)$.

The key idea is to start with a band structure that have both spatial inversion symmetry ($m_z$) and axial spectral symmetry ($2_z$) which allows the existence of a pair of cross points (Dirac points) in 2D Fourier transformed space of hybrid resonant crystals [7,14]. Notice that two symmetrical cross points, with respect to the $\Gamma = 0$ point of the lattice and with different linear polarizations, can be easily obtained in the dispersion curves of hybrid lattice for non-interacting TE and TM waves [1,7]. Moreover, if axial symmetry is removed the two cross points become asymmetric, and, if also spatial inversion is broken, asymmetric gaps open and many different stationary points appears in the non-symmetric dispersion curves. Finally, the time reversal is in any case conserved in these non-magnetic photonic crystals [2,7].

First of all, let us consider un elementary cell, where the exciton transition is turned off, and the optical parameter values of the model are the same of ref.[7], namely: i) $\varepsilon_b = 10.24$ is the background dielectric constant value of the isotropic semiconductor slab, while ii) the optical axis of the uniaxial slab is parallel the x-axis (where $\beta = 0$ is the in-plane angle) and the uniaxial dielectric constant values are chosen as in the ref. [7] namely: $\bar{\varepsilon} = \frac{\varepsilon_{//} + \varepsilon_{\perp}}{2} = \varepsilon_b$ and $\varepsilon_{//} > \varepsilon_{\perp}$ and, for not too large birefringence effect: $\Delta\varepsilon = \varepsilon_{//} - \varepsilon_{\perp} = 4$, the parallel and orthogonal dielectric constants are respectively: $\varepsilon_{//} = 12.24$ and $\varepsilon_{\perp} = 8.24$. iii) The spatial periodicity along z-axis is: $d = \lambda/2 = \frac{\pi c}{\omega_o \sqrt{\varepsilon_b}}$, where $\hbar\omega_o$ is the photon energy resonance of the photonic crystal ($\hbar\omega_o = 1.418\text{eV}$ ref.[7]).

It is well known that, at variance of isotropic photonic crystals [15], the interference between TE and TM linear polarizations in hybrid crystals allows to observe photonic gap inside the Brillouin zone [1,7]. In fact, while the forward and backward waves, with the same polarization, have constructive interference at the high symmetry points of Brillouin zone ($K = 0$ and $K = \pm\pi/d$), the interference between different polarized components can drops inside the Brillouin zone [1,7] giving a strong deformation of the photon density of states [1]. Notice, that the former effect is proportional to the birefringence parameter value $\Delta\varepsilon$ and increases by increasing the photon energy.

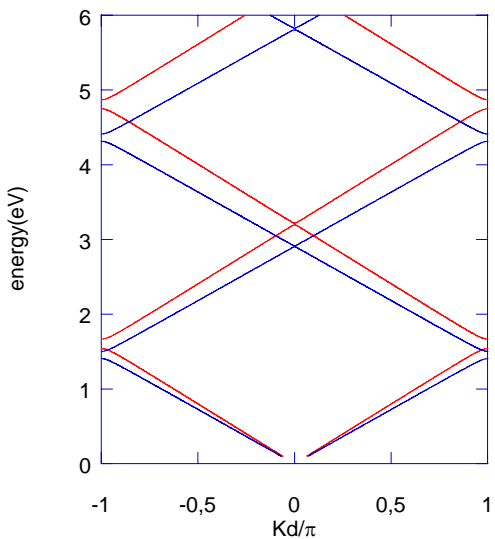
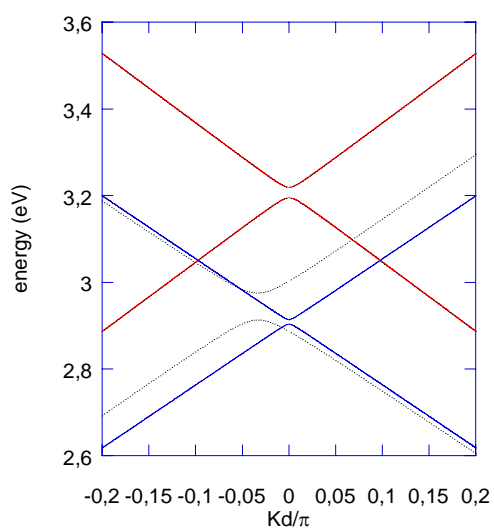

fig.1a  fig.1b

In fig.1a the photon dispersion curves, computed for non-normal propagation ($\alpha = 20^o$), are reported. Notice that two cross points, due to TE (blue) and TM (red) curves at $\hbar\omega \approx 3.07\text{eV}$ photon energy and symmetric with respect to the $\Gamma = 0$ point of the lattice, are clearly shown (see also fig.1b). Moreover, for the in-plane optical axis parallel to the x-axis ($\beta = 0$) with respect to the plane of incidence (x,z), the TM-electric field components (red curves) and the TE-components (blue curves) are not interacting, and many cross points are present close, but not in coincidence [14], with high symmetry points of the lattice.

Finally, no "absolute gaps" along z-direction are shown in the dispersion curves (see fig.1a) [7]. We will use the name "absolute gap in the Brillouin zone" (BZ) in this letter in order to indicate a gap along z-axis with respect to both TE and TM linear polarizations. Notice that, in principle at non-normal incidence the system should not present absolute gap since the in-plane wave vector has a continuum of values.

Now, let us to remove the former spatial symmetry by separating the $\lambda/4$ uniaxial slab in two $\lambda/8$ uniaxial slabs with the two different optical axes performing respectively a $\beta$-angle on (x,y)-plane and a $\gamma$-angle on (x,z)- plane with the x-axis. In conclusion, the total symmetry of the system is driven by three angles: $\alpha, \beta$ and $\gamma$. Moreover, in order to make clear the procedure for the optical tailoring of the sample of our "minimum model", each layer must to take under control only a single optical parameter, and the dielectric tensors of the two uni-axial layers are:

$$\begin{pmatrix} \varepsilon_{//} \cos^2\beta + \varepsilon_\perp \sin^2\beta & (\varepsilon_{//} - \varepsilon_\perp)\sin\beta\cos\beta & 0 \\ (\varepsilon_{//} - \varepsilon_\perp)\sin\beta\cos\beta & \varepsilon_{//} \sin^2\beta + \varepsilon_\perp \cos^2\beta & 0 \\ 0 & 0 & \varepsilon_\perp \end{pmatrix} ; \begin{pmatrix} \varepsilon_{//} \cos^2\gamma + \varepsilon_\perp \sin^2\gamma & 0 & (\varepsilon_{//} - \varepsilon_\perp)\sin\gamma\cos\gamma \\ 0 & \varepsilon_\perp & 0 \\ (\varepsilon_{//} - \varepsilon_\perp)\sin\gamma\cos\gamma & 0 & \varepsilon_{//} \sin^2\gamma + \varepsilon_\perp \cos^2\gamma \end{pmatrix} ,$$

for the optical axes on (x,y)-plane and on (x,z)-plane respectively.

Notice that in this second step of the tailoring, spatial symmetry and axial symmetry will be both removed ($m_z$ and $2_z$) by taking the angle $\gamma \neq n\pi/2$ where $n = 0, \pm 1, \pm 2,...$ [2]. In particular, we have chosen $\gamma = 20°$ in order to maintain a rather large component along the x-axis, while the in-plane optical axis still remain parallel to the x-axis ($\beta = 0$). Let us underline that to remove the axial symmetry gives dispersion curves $\hbar\omega(K_x, K)$ non-symmetric with respect to $\pm K$ Bloch wavevectors ($\hbar\omega(K_x, K) \neq \hbar\omega(K_x, -K)$), but maintains the time inversion $\hbar\omega(K_x, K) = \hbar\omega(-K_x, -K)$ [2].

Finally, the former optical asymmetry is strongly reflected on the symmetry of the transfer matrix and on the properties of the coefficients of the bi-quadratic equation, used for computing the dispersion curves. In fact, in our calculation the electromagnetic ket has the following symmetry: $\begin{pmatrix} E_x \\ H_y \\ E_y \\ H_x \end{pmatrix}$ that, by a smart sum and difference [7], can be reduced to: $\begin{pmatrix} A_x \\ B_x \\ A_y \\ B_y \end{pmatrix}$ where A and B are respectively forward and backward coefficients of the electromagnetic field.

Therefore, the 4X4 transfer matrix is composed by four 2X2 block matrices with respect to the former ket, that, for symmetric dispersion curves with respect to $\pm K$ Bloch wavevectors, has the properties: $T_{1,n}(\omega) = T^*_{2,n}(\omega)$ ; $T_{3,n}(\omega) = T^*_{4,n}(\omega)$ for n=1,2,3,4. (1)

The eigenvalues problem $|\vec{\vec{T}}(\omega) - e^{iKd}\vec{\vec{I}}| = 0$ can be reduced to a bi-quadratic equation solution, for unknowns $Z \equiv e^{iKd}$ : $Z^4 + pZ^3 + qZ^2 + pZ + 1 = 0$ , where the coefficients are real, namely:
$q = |T_{1,1}|^2 + |T_{3,3}|^2 - |T_{1,2}|^2 - |T_{3,4}|^2 + + 2\,\text{Re}(T_{3,3}T_{1,1}) + 2\,\text{Re}(T^*_{3,3}T_{1,1}) - 2\,\text{Re}(T_{3,1}T_{1,3}) - 2\,\text{Re}(T^*_{3,2}T_{1,4})$ ;

$$p = -2\left[\mathrm{Re}(T_{1,1}) + \mathrm{Re}(T_{3,3})\right]. \qquad (2)$$

While for non-symmetric dispersion curves the conditions of eq.1 are verified only for y-components (n=3,4) and the bi-quadratic equation becomes: $Z^4 + pZ^3 + qZ^2 + rZ + s = 0$ with complex coefficients and $s = s' + is''$ is $|s|^2 = 1$. Finally, the solution is obtained by solving the following system of unknowns $Z = X + iY$:

$$\begin{cases} 2aX^2 + 2s''Y^2X^2 + bX + cY + d = 0 \\ 2s''X^2 + 2eYX + fX + gY + h = 0 \end{cases} \text{ and } \begin{cases} a = 1+s'; \ b = p'+r'; \ c = r''-p''; \ d = q'-s'-1 \\ e = 1-s'; \ f = r''+p''; \ g = p'-r'; \ h = q''-s'' \end{cases} \qquad (3).$$

In fig.1b, where the two cross points of the symmetric system of fig.1a are shown in an enlarged energy scale ($\hbar\omega \approx 3.0484\,\mathrm{eV}$ and $Kd \approx \pm 0.0963\,\pi$), we have superimposed the dispersion curves computed by removing axial asymmetry (black curves). It is interesting to underline that ordinary wave (y-components) maintains its symmetry given in eq.(1), while the extra-ordinary wave (x-components) changes in order to move the cross point for K>0 ($\hbar\omega \approx 3.1\,\mathrm{eV}$ and $Kd = 0.0667\,\pi$) and therefore to obtain an asymmetric situation [16].

Finally, for the last step of the sample tailoring, we will have to optimize the β-angle of the in-plane dielectric tensor in order to open a new absolute photonic gap, dues to strong TE-TM interaction [1,7]. In fig.2 a well formed absolute gap (of about ~113mev) between upper and lower photonic bands is computed for in-plane angle $\beta = 60°$, where the lower dispersion curve supports an inflection point with an horizontal axis [2,7].

Notice that the energy difference between the maximum and the inflection points of the lower energy band ($\Delta E_{up} \approx 6\,\mathrm{meV}$) is a direct function of the birefringence parameter ($\Delta\varepsilon = 4$), and gives a limit to the total broadening value of the system in order to observe the oblique unidirectional propagation.

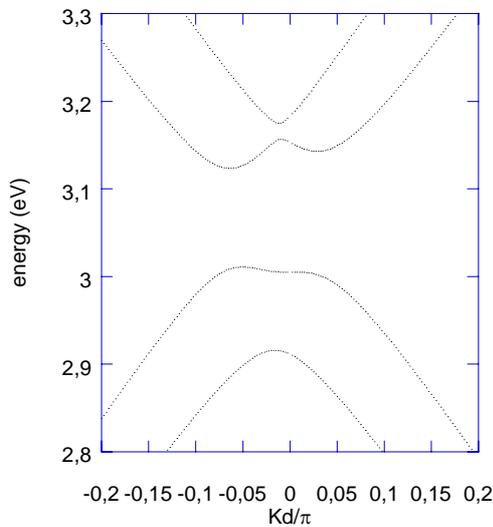
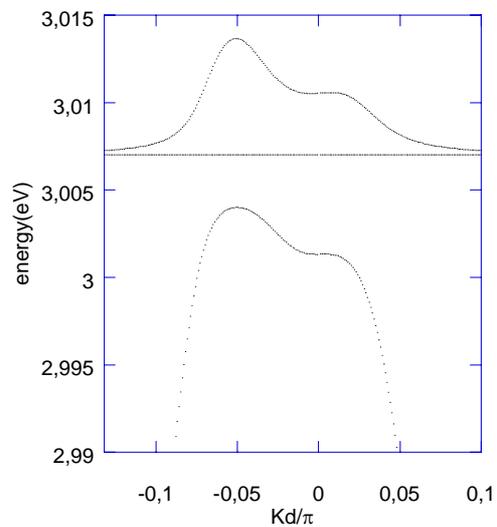

fig.2　　　　　　　　　　　　　　fig.3

A photon energy, in resonance with the inflection point ($\hbar\omega = 3.005eV$), crosses the dispersion curves also in a point with K<0 and group velocity $(\partial\omega_\alpha / \partial K) > 0$, while the inflection point, close to $\Gamma = 0$, has group velocity components: $\vec{v}_g = \left(\frac{c}{\sqrt{\varepsilon_b}\sin(\alpha)}, 0, 0\right)$ that travels along in-plane direction. Therefore, for photon energy in resonance with the inflection point, the system can transport intensities for K<0, while for $K \geq 0$ a stationary wave along z-direction appears, that can sustain a travelling wave propagating along the x-axis.

In conclusion, the former asymmetric absolute gap is obtained, starting from a non-normal totally symmetric ($\beta = \gamma = 0^o$) two-layers sample of fig.1a, by a continuous variation of the optical parameters ($\beta = 60^o; \gamma = 30^o$) that broken spatial and axial symmetry ($m_z, 2_z$), but maintains time reversal symmetry. Since, it is well known [14] that two systems belong to the same topological phase when continuous deformation from one into the other leaves the gaps unchanged, the building up of this new gap should underline the rather different topological properties of the final sample with respect to the starting system of fig.1a.

Now, we would like to study the strong radiation-matter interaction effect (polariton), on the former absolute gap, due to the exciton transition resonance. Therefore let us to turn on the Wannier exciton transition for a direct valence-conduction parabolic band model: $E_{1S}(K_x) = E_{1S}(K_x = 0) + \frac{\hbar^2}{2M}K_x^2$, where $K_x$ is the in-plane wavevector of the exciton centre-of-mass (the total mass is M=0.524 $m_o$). The non-normal optical properties of these resonant non-magnetic photonic crystal, where linear and quadratic spatial dispersion effects are both present, will be computed in the framework of exciton-polariton by self-consistent Maxwell-Schroedinger equation solutions in the effective mass approximation along the line of refs.[3,7], and all the basic formula, necessary for computation, are given in the complementary materials.

The exciton-polariton dispersion curves are computed for exciton energy transition ($E_{1S}(K_x = 0) = 3.007eV$) close to the inflection point of the lower photonic band. In fig.3 the zone of exciton-photon interaction is shown in an enlarged energy scale and two so called "exciton middle-bands" are also observed [3,7].

Notice, that while one middle band is in interaction with the lower photonic band, and reproduces also the lower band shape, the other one, in interaction with the upper photonic band at higher energies ($\hbar\omega \approx 3.107eV$), results rather unperturbed [7]. Moreover, the strong radiation-matter interaction between the former two interacting states is conserved and gives exciton-polariton splitting energy of ~ 8.9meV greater than the broadening of the high quality quantum wells.

Moreover the rather unperturbed middle band introduces a deformation in the dispersion curves that facilitates the presence of an horizontal axis in the correspondence of the inflection point, and this effect should enlarge the dielectric tailoring of these systems in order to obtain oblique unidirectional propagation. At variance of the former unperturbed middle band, the middle band interacting with the lower energy band still remain in the strongest radiation-matter interaction for a rather large range of Bloch wavevector values (about 1/10 of Brillouin zone). Therefore, this effect should give a huge enhancement of the polariton density of states for hybrid resonant crystals and, since it is obtained for no extreme values of the birefringence condition, it is one of the most interesting results of the present work.

In conclusion, unidirectional oblique exciton-polariton propagation in hybrid photonic crystals are theoretically studied in a simple two layer "minimum model" by a self-consistent calculation in the effective mass approximation. The tailoring of the hybrid crystals are performed for not extreme value ($\Delta\varepsilon = 4$) of the birefringence coefficient ($\Delta\varepsilon = \varepsilon_{//} - \varepsilon_{\perp}$), and the optical properties are computed under weak and strong (polaritons) radiation-matter interaction.

Finally, for exciton transition energy in resonance with an inflection point, the exciton-polariton dispersion curves can show a continuum of resonance energies smeared out in a range of the center-of-mass wave vectors as large as 1/10 of the Brilouin zone, and this property is very interesting for fundamental and applications.

Supplementary materials:

Now, let us consider a Wannier exciton transition, between direct valence-conduction electronic bands, where the Kane's energy is: $E_K = 23 eV$ in sound agreement with GaAs/AlGaAs/GaAs(0,0,1) systems [7,8]. Let us consider an "high quality" quantum well [8], where the Wannier exciton is perfectly confined in a slab of thickness: $L_w = 10 nm$, and the non-radiative broadening value is $\Gamma_{NR} \approx 0.25 meV$ [3,8]. The 2D hydrogen variational envelope function is modeled by a simple two sub-bands model:

$$\Psi_{1S}(\vec{r}, Z) = N_{1S} \cos\left(\frac{\pi}{L_w} z_e\right) \cos\left(\frac{\pi}{L_w} z_h\right) e^{-\rho/a_B} e^{iK_x X} / \sqrt{S}, \tag{1}$$

where the normalization coefficient is: $N_{1S} = \frac{1}{L_w a_B} \sqrt{\frac{8}{\pi}}$ and the relative and center-of-mass coordinates are respectively: $\vec{r} = \vec{r}_e - \vec{r}_h = (\vec{\rho}, z)$ and $\vec{R} = \frac{m_e}{M} \vec{r}_e + \frac{m_h}{M} \vec{r}_h = (\vec{R}_{//}, Z)$.

The exciton transition energy value $E_{ex}(K_x = 0)$ is usually computed by variational solution of the Schroedinger equation, where the effective Bohr radius $a_B$ is a minimization parameter [6]. Let us take the exciton parameter values in close resonance with the photon resonance energy of the system ($E_{ex}(K_x = 0) = 1.418 eV \approx \hbar\omega_o$ and the effective Bohr radius is: $a_B = 8.047 nm$). Notice that in the present computation, in order to tune the exciton transition in resonance with higher energy bands of the system, we will change the $E_{ex}(K_x = 0)$-value only, while all the further exciton parameters are taken unchanged [7]. The exciton-polariton non-local dielectric function is:

$$\varepsilon(Z, Z') = \varepsilon_w \delta(Z - Z') + 4\pi g \frac{S_{ex}(\omega) \Psi^*_{1S}(\vec{r}=0, Z) \Psi^*_{1S}(\vec{r}=0, Z')}{\left(E_{1S}(K_x)\right)^2 - \left(\hbar\omega\right)^2 - 2i\hbar\omega\Gamma_{NR}} \tag{2}$$

where g is spin-orbit coefficient and $\Gamma_{NR}$ is the non-radiative broadening, and $S_{ex}(\omega) = \frac{E_K}{\hbar\omega} \frac{\hbar^2 e^2}{m_o}$. The complex radiation-matter self-energy is given by:

$$\Sigma_\eta(\omega) = 4\pi g\, q_\eta^2\, S_{ex}^{RWA}(\omega)\, M_{ex}(\omega) \qquad \text{for } \eta = x, y \tag{3}$$

where $S_{ex}^{RWA}(\omega) = \frac{S_{ex}(\omega)}{2 E_{ex}(K_x)}$ and $q_y = \frac{\omega}{c}$, $q_x = \frac{k_w}{\sqrt{\varepsilon_w}}$ and $k_w = \left[\frac{\omega^2}{c^2} \varepsilon_w - K_x^2\right]^{1/2}$.

The complex matrix element $M_{ex}(\omega)$ is:

$$M_{ex}(\omega) = \frac{N_{ex}^2}{4} \left\{ L_w \left[\frac{1}{k_w^2} - \frac{1}{2} \frac{1}{k_{ex}^2 - k_w^2}\right] + i k_w \left[\frac{1}{k_w^2} + \frac{1}{k_{ex}^2 - k_w^2}\right]^2 \left[e^{ik_w L_w} - 1\right] \right\}, \tag{4}$$

where $k_{ex} = \frac{2\pi}{L_w}$. The dispersion curves are computed for g=1 and: $\Gamma_{NR} \to 0$.